\documentstyle[12pt,psfig,epsf]{article}
\begin{document}
\baselineskip 0.8cm
\title{Spin effects in single-electron tunneling \\ in magnetic junctions }
\author{J. Martinek$^1$, J.Barna\'s$^{2,3}$, G. Micha\l ek$^1$,
B.R. Bu\l ka$^1$ and A. Fert$^3$ \\
$^1$ Institute of Molecular Physics,
     Polish Academy of Sciences,  \\  
     ul. Smoluchowskiego 17, 60-179 Pozna\'n, Poland \\
$^2$ Department of Physics, A. Mickiewicz University, \\
     ul. Umultowska 85, 61-614 Pozna\'n, Poland \\
$^3$ Unite Mixte de Physique CNRS/Thomson, \\ Domaine de Corbeville,
     91-404 Orsay, France }

\maketitle

\noindent

Spin dependent single electron tunneling  in ferromagnetic
double junctions is analysed theoretically in the limit of sequential
tunneling. The influence of discrete energy spectrum of 
the central electrode (island)
on the spin accumulation, spin fluctuations and
tunnel magnetoresistance is analysed numerically in the
case of a nonmagnetic island. 
It is shown that spin fluctuations are significant 
in magnetic as well as in nonmagnetic junctions. 

\newpage

Single electron tunneling 
in mesoscopic double-junctions has been recently  
extensively studied both experimentally and theoretically.$^{1-3}$  
For a small capacitance $C$ of the central electrode (island)
the charging energy $E_c=e^2/2C$ can be
larger than the thermal energy $k_BT$. 
Discrete charging of the island leads then to Coulomb
blockade of the electric current at 
voltages below a certain
threshold voltage, and to a characteristic 
'Coulomb staircase' at higher voltages.
When the number of electrons on the island
is not large, the energy quantization may become important as well,
and can lead to additional steps in the $I-V$ curves (or additional peaks in
the $dI/dV$ characteristics).$^{4,5}$

The interplay of ferromagnetism and discrete charging was studied only
very recently.$^{6-11}$ Theoretical results show  
that discrete charging
leads to oscillations in tunnel magnetoresistance (TMR) 
(that is in the change of the total junction 
resistance when relative orientation of the magnetization of
both electrodes and of the island is varied).$^{9,11}$ 
In Ref.[9] the intrinsic spin relaxation time
on the island was assumed to be sufficiently short
(shorter than the time between two successive tunneling events)
to neglect spin accumulation.  This model has then been extended
to describe the regime where intrinsic spin relaxation time on the island
is larger than the time between successive tunneling events. In this regime
spin accumulation has to be taken into account, which leads to enhanced TMR
and several new phenomena, like inverse TMR or  negative differential
resistance. Spin accumulation can also generate TMR when the island is
nonmagnetic.

In the model of Refs.[9,11] the discrete nature of the density of
states on the island and also the fluctuations of the spin accumulation
were neglected.
These restrictions are relaxed in the
present paper, where spin fluctuations as well as discrete
structure of the density of single-electron states are taken into account.
On the other hand, as in the models refered to above, we restrict 
our considerations to the limit of sequential
tunneling (where the {\it orthodox} tunneling theory is applicable),
that is to the limit
when the resistance of each single junction is larger than the quantum
resistance $R_q$, $R_q=h/e^2$. In that limit higher order processes
(cotunneling) can be neglected. However, these processes
may play an important role in the blockade regime.$^{10}$

Geometry of the junction and
energy structure is shown schematically in Fig.1. In a general case
both electrodes and the island can be ferromagnetic. 
We consider the case when the resistances of 
both left and right junctions are much larger than the quantum
resistance $R_q$. Moreover, we assume that $k_BT>>\Gamma$, where
$\Gamma$ denotes width of the discrete energy levels on the island.
To calculate the electric current $I$ which flows through the junction  
when a voltage $V$ is applied between the source (right)
and drain (left) electrodes,
one can then use the {\it orthodox} tunneling theory.$^{4,5,12-14}$

Assume that there are $N^*$ excess electrons on the island.
Generally, one can write
$N^*=N^*_\uparrow +N^*_\downarrow$, where  $N^*_\sigma$
($\sigma =\uparrow$ or $\sigma =\downarrow$) is the number of
excess electrons of a given spin orientation.
The excess spin polarization of the island is then equal to 
$N^*_\uparrow -N^*_\downarrow$.
If $N_\sigma$ is the number of electrons with spin $\sigma$ 
on the island when a voltage $V$ is applied, and 
$N^0_\sigma$ is this number at $V=0$, 
then one can write $N^*_\sigma=N_\sigma -N^0_\sigma$. 

Denoting by $E_{i\sigma}$ the discrete energy levels of the island, one
can express the electric current $I$ in the stationary state as
\begin{displaymath}
I=e \sum_{\sigma} \sum_{i} \sum_{\{n\}}
  |T^l_{i\sigma}|^2 P(\{n\})
\end{displaymath}
\begin{equation}
\times \left\{ \delta_{n_{i\sigma},1} [1-f(E_{i\sigma}+E^{l-}_{N^*}-E_F)]
 -\delta_{n_{i\sigma},0} f(E_{i\sigma}+E^{l+}_{N^*}-E_F) \right\}\;.
\end{equation}
In the above equation 
$P(\{n\})$ is the probability of
a particular configuration of
the occupation numbers of discrete energy levels of the island, 
$\{n\}\equiv
\{n_{1\uparrow},n_{2\uparrow},....,n_{1\downarrow},
n_{2\downarrow},.....\}$, 
with 
$n_{i\sigma}=1$ ($n_{i\sigma}=0$) for occupied (empty) states
and the sum over $\{n\}$ denotes the summation over
all possible occupation configurations. 
This probability can be determined
from an appropriate master equation.$^{4,5}$
Apart from this, $T^l_{i\sigma}$ 
in Eq.(1) is the matrix element corresponding to the electron tunneling
to (from) the left electrode from (to) the
level $E_{i\sigma}$ of the island, while $E^{l\pm}_{N^*}$ is defined as 
$E^{l\pm}_{N^*}=eV^{l}_{N^*}\pm E_c$. Here,  
$V^l_{N^*}=(C_r/C)V+N^*e/C$, 
where $C_r$ is the capacitance of the right 
junction,  $C$
is the total capacitance of the island, and $-e$ is the
electron charge ($e>0$).
Finally, we assumed in Eq.(1) the 
equilibrium Fermi-Dirac distribution of
electrons in the electrodes, with the Fermi level  $E_F$ at  
vanishing voltage.
In the following, we will also introduce the quantity
$E^{r\pm}_{N^*}=eV^r_{N^*}\mp E_c$, where  
$V^r_{N^*}=(C_l/C)V-N^*e/C$ and $C_l$ is the capacitance of the
left junction ($C=C_l+C_r$), and the matrix element $T^r_{i\sigma}$ 
for tunneling through the right junction.

It is convenient to
introduce the probability $P(N_{\uparrow},N_{\downarrow})$
of finding on the island $N_{\uparrow}$ electrons with spin
$\sigma =\uparrow$ and $N_\downarrow$ electrons with 
spin $\sigma=\downarrow$,
\begin{equation}
P(N_{\uparrow},N_{\downarrow})= \sum_{\{n\}} P(\{n\})
\delta_{N_{\uparrow},\sum_{p}n_{p\uparrow}} \;
\delta_{N_{\downarrow},\sum_{p} n_{p\downarrow}}\;,
\end{equation}
and the distribution function 
\begin{equation}
F(E_{i\sigma}|N_{\uparrow},N_{\downarrow})
=\frac{1}{P(N_{\uparrow},N_{\downarrow})} \sum_{\{n\}} P(\{n\})
\delta_{n_{i\sigma},1}\;\delta_{N_{\uparrow},\sum_{p}n_{p\uparrow}}
\;\delta_{N_{\downarrow},\sum_{p} n_{p\downarrow}}.
\end{equation}
Equation (3) describes
the probability that the level $E_{i\sigma}$ is occupied under
the condition that the island contains $N_{\uparrow}$ electrons with
spin $\sigma =\uparrow$ and $N_{\downarrow}$ electrons with
$\sigma =\downarrow$.
The current $I$ is then given by
\begin{displaymath}
I=-e\sum_{N_\uparrow}\sum_{N_\downarrow}
\sum_{\sigma}\sum_{i}
P(N_{\uparrow},N_{\downarrow}) \Bigl\{ [1-F(E_{i\sigma}|N_{\uparrow},
N_{\downarrow})]|T^l_{i\sigma}|^2 f(E_{i\sigma}+E^{l+}_{N^*}-E_F)
\end{displaymath}
\begin{equation}
 - F(E_{i\sigma}|N_{\uparrow},N_{\downarrow}) |T^l_{i\sigma}
 |^2 [1-f(E_{i\sigma}+E^{l-}_{N^*}-E_F)] \Bigr\} \;.
\end{equation}

In the stationary state one finds the following master equation for the 
probability $P(N_{\uparrow},N_{\downarrow})$:
\begin{displaymath}
{\partial \over \partial t} P(N_{\uparrow},N_{\downarrow})=
0=-P(N_{\uparrow},N_{\downarrow})A(N_{\uparrow},N_{\downarrow})
\end{displaymath}
\begin{displaymath}
+P(N_{\uparrow}+1,N_{\downarrow}) B_{\uparrow}(N_{\uparrow}
+1,N_{\downarrow}) 
+P(N_{\uparrow},N_{\downarrow}+1) B_{\downarrow}
(N_{\uparrow},N_{\downarrow}+1)
\end{displaymath}
\begin{equation}
+ P(N_{\uparrow}-1,N_{\downarrow})C_{\uparrow}(N_{\uparrow}-1,
N_{\downarrow}) 
+ P(N_{\uparrow},N_{\downarrow}-1)C_{\downarrow}
(N_{\uparrow},N_{\downarrow}-1) + {\cal R}_s,
\end{equation}
where 
\begin{displaymath}
A(N_{\uparrow},N_{\downarrow})=\sum_{\sigma}\sum_{i}
[1-F(E_{i\sigma}|N_{\uparrow},N_{\downarrow})]
\{|T^l_{i\sigma}|^2 f(E_{i\sigma}+E^{l+}_{N^*}-E_F)
\end{displaymath}
\begin{displaymath}
+|T^r_{i\sigma}|^2 f(E_{i\sigma}-E^{r+}_{N^*}-E_F) \}
+\sum_{\sigma}\sum_{i}F(E_{i\sigma}
 |N_{\uparrow},N_{\downarrow})\{|T^l_{i\sigma}|^2
[1-f(E_{i\sigma}+E^{l-}_{N^*}-E_F)]
\end{displaymath}
\begin{equation}
+ |T^r_{i\sigma}|^2[1-f(E_{i\sigma}
-E^{r-}_{N^*}-E_F)] \} \;
\end{equation}
\begin{displaymath}
B_{\sigma}(N_{\uparrow},N_{\downarrow})=
\sum_{i}F(E_{i\sigma}|N_{\uparrow},N_{\downarrow})
\{|T^l_{i\sigma}|^2 [1-f(E_{i\sigma}+E^{l-}_{N^*}-E_F)]
\end{displaymath}
\begin{equation}
+|T^r_{i\sigma}|^2[1-f(E_{i\sigma}-E^{r-}_{N^*}-E_F)] \} \;
\end{equation}
\begin{displaymath}
C_{\sigma}(N_{\uparrow},N_{\downarrow})= \sum_{i}[1-F(E_{i\sigma}
|N_{\uparrow},N_{\downarrow})] \{|T^l_{i\sigma}|^2 f(E_{i\sigma}
+E^{l+}_{N^*}-E_F)
\end{displaymath}
\begin{equation}
+|T^r_{i\sigma}|^2 f(E_{i\sigma}-E^{r+}_{N^*}-E_F) \}\;.
\end{equation}
In Eq.(5) ${\cal R}_s$ stands for terms responsible for magnetic relaxation
on the island.
Note that spin-conserving
relaxation processes are included in Eq.(5) only
through the distribution function
$F(E_{i\sigma}|N_\uparrow N_\downarrow)$.

We assume in the following
that the energy relaxation time
due to spin-conserving relaxation processes
is significantly shorter than the time
between successive tunneling events, and also shorter
than the spin relaxation time. This allows us to use an equilibrium 
form $F_{\rm eq}(E_{i\sigma}|N_\uparrow N_\downarrow)$ 
of the distribution function $F(E_{i\sigma}|N_\uparrow N_\downarrow)$,
which depends on the spin orientation.
In a general case the equilibrium distribution 
$F_{\rm eq}(E_{i\sigma}|N_\uparrow N_\downarrow)$
can be found from the Gibbs distribution. 
In the limit $k_BT \gg\Delta E$ the distribution function
$F_{\rm eq}(E_{i\sigma}|N_\uparrow N_\downarrow)$
is equal to   
by the Fermi-Dirac distribution
\begin{equation}
F(E_{i\sigma}|N_{\uparrow},N_{\downarrow})
=f(E_{i\sigma}-\mu_\sigma (N_{\sigma}))\; ,
\end{equation}
where the chemical potential $\mu_\sigma (N_{\sigma})$ is 
determined by the equation
\begin{equation}
\sum_{i}f(E_{i\sigma} -\mu_\sigma (N_{\sigma}))=N_{\sigma}\; .
\end{equation}
As noticed by Beenakker [5], the equilibrium distribution
for $k_BT\approx\Delta E$ differs significantly from the Fermi
distribution. 

Consider now some numerical results. For simplicity, we will restrict
the following considerations to the
situation when the island is nonmagnetic
and the magnetizations of the electrodes can be either
parallel or antiparallel. To emphasize the role of
spin accumulation we assume that
the intrinsic spin relaxation time on the island is long enough 
to neglect all intrinsic spin-flip 
processes. The magnetic relaxation on the island takes then place
only through tunneling processes.
For simplicity, we assume that the energy levels
on the island are equidistinct, with the
inter-level spacing $\Delta E$. Apart from this, we assume that 
$k_BT<\Delta E$ (but still $k_BT>>\Gamma$) 
and $\Delta E<E_c$.
In Fig.2a we show the I-V characteristics for parallel and
antiparallel configurations.
In both cases the electric current is blocked
below a threshold voltage equal approximately to 13 mV. Above
the threshold voltage typical 'Coulomb staircase' appears with
additional small steps due to discrete levels. 
The existence of those steps leads to 
additional peaks in the derivative $dI/dV$,
as shown in Fig.3b for the antiparallel
configuration (the corresponding curve for the parallel configuration
was not shown there for clarity).

The difference between I-V curves for the parallel and antiparallel
configurations is due to a different spin accumulation in both geometries
(note, that both I-V curves become identical
when no spin accumulation occurs in the nonmagnetic island$^{9,11}$).
To show correlations between the I-V curves and spin accumulation,
we present in Fig.2c the average value of the difference
between the numbers of spin-up and spin-down excess electrons on the island,
$\langle N^*_\uparrow -N^*_\downarrow \rangle$,
for both antiparallel and parallel
configurations (up to a constant factor, Fig.2c shows
an average magnetic moment
induced on the island due to spin accumulation).
For the symmetrical structure assumed here
($R_{l\uparrow}/R_{l\downarrow}=R_{r\uparrow}/R_{r\downarrow}$ for the parallel
configuration, where $R_{l(r)\sigma}$ are the junction resistances for a
given spin channel) 
there is no significant
spin accumulation
in the parallel configuration.
Only small nonzero values occur 
around the Coulomb steps. For asymmetrical junctions spin accumulation 
also occurs in the parallel configuration. Discrete structure
of the density of states
on the island appears in the voltage dependence of
$\langle N^*_\uparrow -N^*_\downarrow \rangle$ 
as the fine steps clearly visible in Fig.2c. 

The number 
$N^*_\uparrow -N^*_\downarrow$,
of spins accumulated on the island
fluctuates around its average value 
$\langle N^*_\uparrow -N^*_\downarrow \rangle$,
as shown in Fig.2d,
where the standard deviation
$[\langle (N^*_\uparrow -N^*_\downarrow )^2 \rangle -
\langle N^*_\uparrow -N^*_\downarrow \rangle^2]^{1/2}$
is plotted against the voltage $V$.
It is worth to note that although there  is almost no spin accumulation
in the parallel configuration, the corresponding
fluctuations are relatively large and
comparable with the fluctuations in the antiparallel configuration.
The fine structure in the voltage dependence of the standard
deviation originates from the discrete electronic structure of the island.

As we have
already pointed above, spin accumulation on the island gives rise 
to a difference between the I-V curves in the parallel and antiparallel
configurations. This difference, in turn, leads
to the tunnel magnetoresistance
(TMR) defined quantitatively 
as $(R_{ap}-R_p) /R_p$, where $R_{ap}$
and $R_p$ are the total
junction resistances respectively in the antiparallel
and parallel configurations. The ratio $(R_{ap}-R_p) /R_p$
is shown in Fig.2e, where the broad peaks
correspond to the Coulomb steps while the fine structure originates from 
the discrete structure of the density
of states.

In the limit of a nonmagnetic junction (nonmagnetic island and
nonmagnetic electrodes) there is no spin accumulation on the island
and no TMR. However, spin fluctuations 
still occur as shown in Fig.3 for the limit of no intrinsic spin
relaxation on the island. As in Fig.2, the discrete energy levels
on the island lead
to fine steps in the voltage dependence of the standard deviation. 

In summary, we developed in this letter  a formalism
for calculating electric current and tunnel
magnetoresistance in ferromagnetic
double junctions, which takes into account
spin accumulation on the island and 
discrete structure of the density of states. 
The discrete energy levels lead to fine steps in the  $I-V$ curves and to additional
peaks in the first derivative $dI/dV$. Moreover,
we showed that the discrete levels give rise to fine structure in
the spin accumulation and spin fluctuations on the
island, as well as in TMR.
To single out the charging effects we neglected
in this letter tunneling processes with simultaneous spin flip, which
lower TMR and also 
lead to a smoth decrease of TMR with increasing voltage.$^{15,16}$

The work was supported by the Polish Research Committee
through the research  project 2 P03B 075 14.

\pagebreak

{\bf References}
\vskip 0.5cm
\begin{enumerate}
\item {D.V. Averin and K.K. Likharev, J. Low Temp. Phys.,
     {\bf 62}, 345  (1986).}
\item {{\it Single Charge Tunneling},
  editted by H. Grabert and M.H. Devoret, NATO ASI
  Series vol 294 (Plenum Press, New York 1992).}
\item {G. Sch\"on, in {\it Quantum Transport and Dissipation}, eds.
      T. Dittrich, P. H\"anggi, G.-L. Ingold, B. Kramer, G. Sch\"on, and
      W. Zwerger, Wiley - VCH Verlag, 1998, pp 149-212}
\item {D.V. Averin and A.N. Korotkov, Zh. Eksp. Teor. Fiz. {\bf97},
 1661 (1990).}
\item {C.W.Beenakker, Phys. Rev. B{\bf44}, 1646 (1991).}
\item {J. Inoue and S. Maekawa, Phys. Rev. B {\bf 53}, R11927 (1996).}
\item {K. Ono, H. Shimada, S. Kobayashi and Y. Outuka,
      J. Phys. Soc. Japan {\bf 65}, 3449 (1996).}
\item {L.F. Schelp, A. Fert, F. Fettar, P. Holody, S.F. Lee, J.L. Maurice,
  F. Petroff and A. Vaures,  Phys. Rev. B {\bf 56}, R5747  (1997).}
\item {J. Barna\'s and A. Fert, Phys. Rev. Lett. {\bf 80}, 1058 (1998).}
\item {S. Takahashi and S. Maekawa, Phys. Rev. Lett. {\bf 80},  (1998).} 
\item {J. Barna\'s and A. Fert, submitted}
\item {M. Amman, R. Wilkins, E. Ben-Jacob, P.D. Maker and
  R.C. Jaklewic, Phys. Rev. B  {\bf 43}, 1146  (1991).}
\item{ D.C. Ralph, C.T. Black and M. Tinkham,
      Phys. Rev. Lett. {\bf 74}, 3241 (1995).}
\item{D. Porath, Y. Levi, M. Tarabiah and O. Millo, 
      Phys. Rev. B  {\bf 56}, 9829 (1997).}
\item {J.S Moodera, L.R. Kinder, T.M. Wong and R. Meservey,
    Phys. Rev. Lett. {\bf 74}, 3273  (1995).}
\item {S. Zhang, P.M. Levy, A.C. Marley and S.S.P Parkin,
    Phys. Rev. Lett. {\bf 79}, 3744  (1997).}

\end{enumerate}
\vskip 3cm

\newpage

\begin{description}
\begin{figure}
\epsfxsize=14cm
\epsfbox{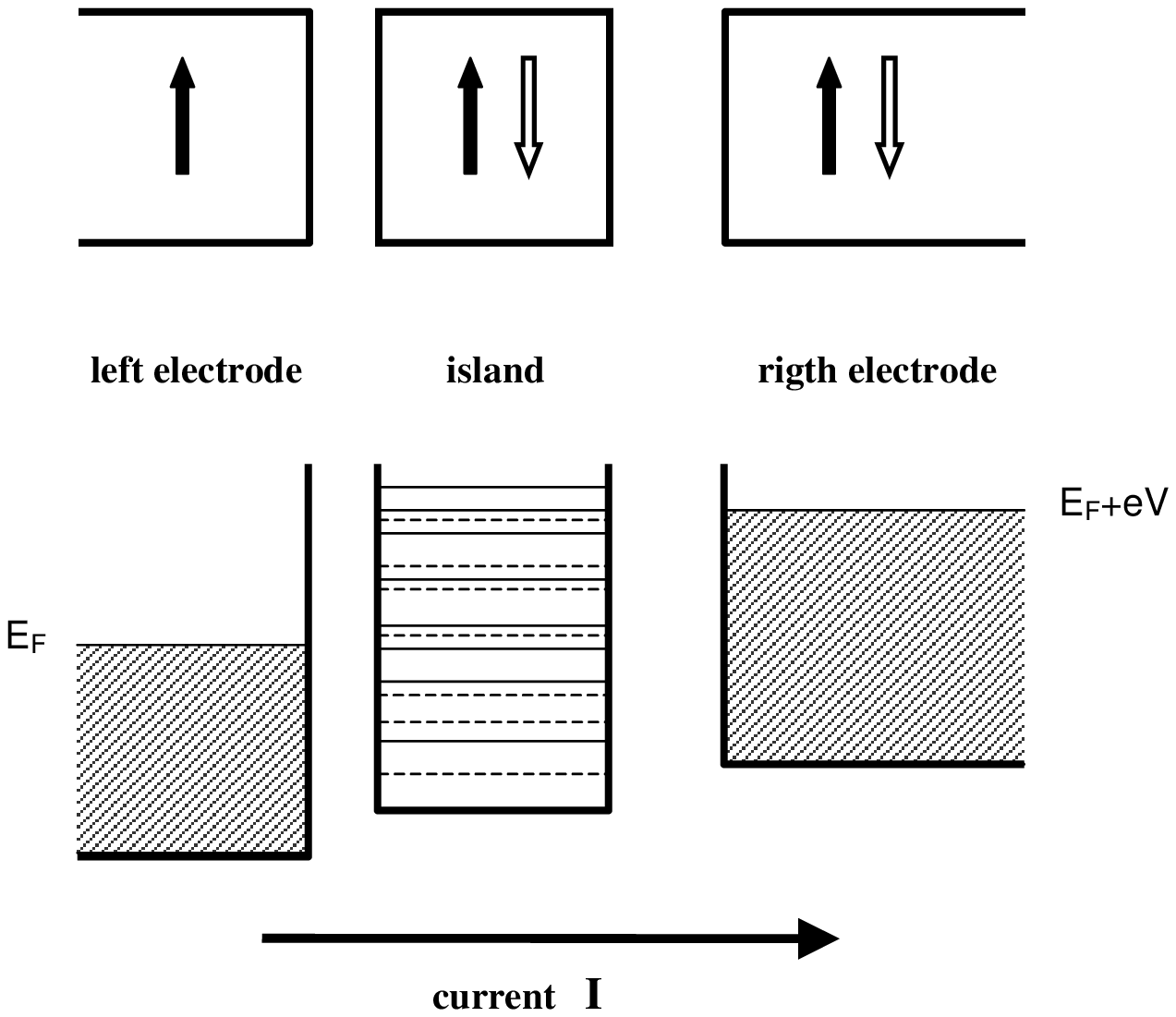} 

\item [Fig.1]
     Schematic of the double-junction considered in this paper.
     Magnetization of the island and of the right electrode can be either
     parallel or antiparallel to the magnetization of the left electrode,
     as indicated by the solid and open arrows. Discrete levels
     of the island are marked schematically by the solid
     (dashed) lines for spin-up (spin-down) 
     electron states. 
\end{figure}
\begin{figure}
\epsfxsize=6.8cm
\epsfbox{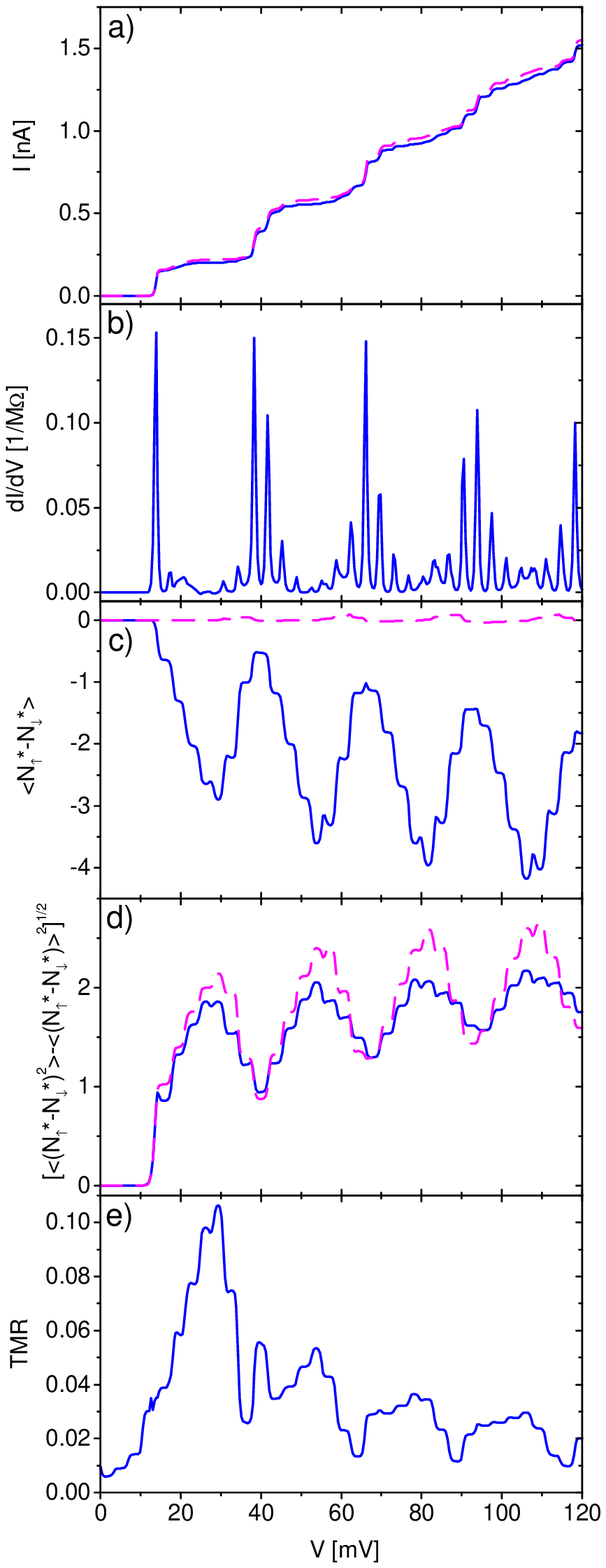} 
\item [Fig.2]
   Voltage dependence of the tunnel current $I$ (a), first derivative
   $dI/dV$ (b), spin
   accumulation
   $\langle N^*_\uparrow -N^*_\downarrow \rangle$
   (c),  standard
   deviation
   $[\langle (N^*_\uparrow -N^*_\downarrow )^2 \rangle -
   \langle N^*_\uparrow -N^*_\downarrow \rangle^2]^{1/2}$
   (d), and tunnel magnetoresistance
   TMR (e), calculated at $T=2.3$K.
   The solid and dotted curves in (a), (c) and (d)
   corespond to the antiparallel and
   parallel configurations, respectively. 
   The other parameters assumed in the numerical calculations are: 
   $\Delta E=3$meV, $C_1=6.6$aF, $C_2=1.32$aF
   (which gives $E_c=10.1$meV), $R_{l\uparrow}=200$M$\Omega$,
   $R_{l\downarrow}=100$M$\Omega$,
   $R_{r\downarrow}=4$M$\Omega$ and $R_{r\uparrow}=2$M$\Omega$
   for the parallel configuration
   ($R_{r\downarrow}=2$M$\Omega$, $R_{r\uparrow}=4$M$\Omega$
   for the antiparallel configuration).
\end{figure}
\begin{figure}
\epsfxsize=14cm
\epsfbox{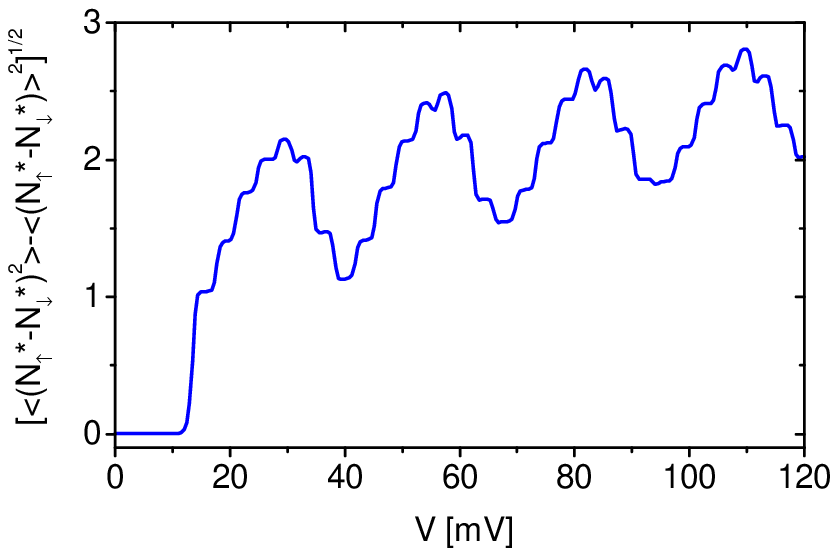} 
\item [Fig.3] Fluctuations of the spin accumulated
   on the island
  in the limit of a nonmagnetic double junction. Standard
  deviation
   $[\langle (N^*_\uparrow -N^*_\downarrow )^2 \rangle -
   \langle N^*_\uparrow -N^*_\downarrow \rangle^2]^{1/2}$
   is shown as a function of the voltage
   $V$ for the parameters: $T=2.3$K, 
  $\Delta E=3$meV, $C_1=6.6$aF, $C_2=1.32$aF
  ($E_c=10.1$meV), $R_{l\uparrow}=R_{l\downarrow}=100$M$\Omega$,
  and $R_{r\uparrow}=R_{r\downarrow}=4$M$\Omega$.
\end{figure}

\end{description}

\end{document}